\documentclass[11pt,twoside]{article}
\usepackage{asp2010}

\resetcounters

\begin{document}

\title{The impact of mergers in the mass distribution of white dwarfs}
\author{J. Isern,$^{1,2}$ 
        S. Catal\'an,$^3$ 
        E. Garc\'{\i}a--Berro,$^{4,2}$, and 
        M. Hernanz$^{1,2}$}
\affil{$^1$Institut de Ci\`encies de l'Espai (CSIC), 
           Facultat de Ci\`encies, Campus UAB, 
           Torre C5--parell, 
           08193 Bellaterra (Barcelona), 
           Spain}
\affil{$^2$Institute for Space Studies of Catalonia (IEEC), 
           c/ Gran Capit\`a 2--4, Edif. Nexus 104, 
           08034 Barcelona, 
           Spain}
\affil{$^3$Centre for Astrophysics Research, 
           University of Hertfordshire, 
           College Lane, Hatfield AL10 AB, 
           United Kingdom}
\affil{$^4$Departament de F\'{\i}sica Aplicada, 
           Universitat Polit\`ecnica de Catalunya, 
           c/ Esteve Terrades 5, 
           08860 Castelldefels, 
           Spain}

\begin{abstract}
Recent  surveys   have  allowed  to   derive  the  white   dwarf  mass
distribution  with  reasonable accuracy.   This  distribution shows  a
noticeable  degree of  structure that  it is  often attributed  to the
evolution of close binaries in  general, and to mergers in particular.
To analyze  if the origin  of this structure  can be attributed  to the
merger of  double white dwarfs,  we have used a  simplified population
synthesis model  that retains the essential processes  of formation of
double  degenerate binaries.  Special  care has  been  taken to  avoid
artifacts introduced by discontinuities in the distribution functions.
Our result is  that these structures are not  probably due to mergers,
but they can  provide a deep insight on the  evolution of close binary
systems.
\end{abstract}

\section{Introduction}

White dwarfs are the most common stellar remnants of the Galaxy, and a
good fraction of them are members of binary systems. A recent study of
the local neighborhood \citep{hol08} has shown that the $\sim 25\%$ of
the white dwarfs  within 20~pc from the Sun are  in binary systems and
that  approximately $\sim  6\%$ of  them are  double white  dwarfs (DD
systems). Because of the emission of gravitational waves, close enough
degenerate pairs loose angular momentum and eventually will merge.

The process  of merging has been  modeled by several  groups using SPH
techniques     \citep{benz90,rasi95,segr97,guer04,yoon07,lore09}    or
grid-based techniques \citep{dsou06,motl07}.  Both sets of simulations
suggest that there is not a prompt explosion after the coalescense and
that a rotating hot corona surrounded  by a thick disk form around the
most  massive white  dwarf.  The  amount  of matter  lost during  this
process is  very small \citep{guer04} but the  subsequent evolution of
this structure is rather  uncertain due to the difficulties introduced in modelling
by the different time scales involved in the process. The key issue in
determining the outcome  of the interaction is the  rate and the total
mass  accreted by  the primary  after  the merger,  and this  strongly
depends on  the coupling between the  rotating star, the  disk and the
magnetic fields  that contribute to the transport  of angular momentum
\citep{pier03b,pier03a,saio04,shen12}

It is generally accepted that if  the merging white dwarfs are made of
carbon  and oxygen, their  total mass  is larger  than Chandrasekhar's
mass, and no  mass is ejected from the system,  the final object would
eventually collapse  to form a neutron star  \citep{nomo91} or explode
as a  Type Ia  supernova \citep{webb84,iben84}. If  the final  mass is
smaller than  the critical mass,  either because the mass  losses from
the  disk are  large enough,  or because  the initial  total  mass was
initially  smaller, the  final  outcome  is a  new,  more massive  and
probably  peculiar white  dwarf.  This  scenario has  been  invoked to
account for the existence of  bright massive white dwarfs in the halo,
the  presence  of dusty  disks  around  white  dwarfs with  metal-rich
atmospheres  \citep{garc07},   to  explain  R~Corona   Borealis  stars
\citep{webb84,long11}    or   strong-field    magnetic    white   dwarfs
\citep{garc12}. In  the case  of the merging  of a  CO and a  He white
dwarf, the outcome can be a thermonuclear explosion \citep{nomo82b} or
a white  dwarf with  a mass  equal or smaller  than the  total initial
mass. Finally, if both white dwarfs  are made of He, the final outcome
would be  the destruction  of the  object, or the  formation of  a new
white dwarf.

At  present, there  are several  deep surveys  that have  provided the
necessary  data to compute  reliable white  dwarf mass  functions: the
Palomar  Green survey  \citep{lieb05}, the  SDSS \citep{kepl07,dege08}
and   the    recent   analysis   of   the    solar   neighborhood   of
\cite{giam12}. These  mass functions display some  degree of structure
as compared with what should  be expected from the evolution of single
white dwarfs.  Besides the  typical main peak \citep{koes79}, the most
important features are the presence  of a low-mass population --- that
is usually attributed to the evolution of close binaries --- an excess
of  stars with  masses ranging  from 0.8  to $1.1\,  M_{\sun}$,  and a
possible  peak around  $1.2\, M_{\sun}$.   These features  are usually
attributed to  white dwarf  mergers \citep{mars97}, although  they are
strongly affected by selection effects \citep{lieb05}.

\section{Method of calculation}

In this work we use a simple  toy model to follow the evolution of the
populations of single and binary white dwarfs. Our simulations aim
to  predict the  expected  number of  DD  mergers \citep{iser97}.   To
compute the distribution of  double degenerate binaries it is necessay
to use a code able to examine,  case by case, the final outcome of all
the initial configurations. Since we are only interested in the impact
of the mergers of DD systems on the mass distribution of white dwarfs,
we have  only considered the  situations in which Roche  lobe overflow
occurs when the  envelope of the star is  convective and when magnetic
braking is effective.
 
The evolution of  a binary system which experiences  a common envelope
phase is  still plagued  with uncertainties. Here  we assume  that the
orbital  energy  is  invested   in  evaporating  the  common  envelope
\citep{iben84} in such a way  that the new separation after the common
envelope phase is given by:
\begin{equation}
A' = \alpha \frac{{M_{1{\rm r}} M_2 }}{{M_1^2 }}
\end{equation}
where $\alpha$  is a dimensionless  free parameter that  describes the
efficiency of the  energy transfer, $M_{1{\rm r}}$ is  the mass of the
remnant, $M_1$ is the original mass  of the primary, $M_2$ is the mass
of the secondary and $A$ is the initial separation. Since the value of
$\alpha$  is  not  known,  we  have adopted  $\alpha  =  1$,  although
different   values   and  functional   dependences   have  also   been
investigated.

Concerning the collective properties  of binaries necessary to perform
the calculations, we  have adopted the following ones:  i) The initial
mass function is written as the mass distribution of the primary times
the   mass   ratio  distribution   \citep{yung93},   while  the   mass
distribution of  the primary star is  taken to be  a simple Salpeter's
distribution in the range  $0.1\leq M_1/M_{\sun}\leq 100$ and the mass
ratio distribution as $f(q)\propto q^n$, where $q=M_2/M_1$ with $n=1$.
ii)   The   adopted  distribution   of   separations   is  $   H(A_0)=
(1/5)\log(R_{\sun}/A_0)$  and,  in order  to  maximize  the impact  of
mergers  in the  mass distribution  function, we  assumed  that single
white dwarfs are well represented by the distribution of wide binaries
\citep{yung93}.   iii) It is  also assumed  a constant  star formation
rate, although  several exponentially-decreasing formation  rates with
different time  scales have also been  considered. iv) The  age of the
Galactic disk is taken  to be 10.5 Gyr, but values as  high as 13 have
also been considered. v) The influence  of metallicity on the age of the
progenitors  and  the mass  of  the  resulting  white dwarf  has  been
neglected, and it  is assumed that all white  dwarfs more massive than
$1.05\, M_{\sun}$ are made of oxygen and neon. vi) We have also assumed that all
the binary systems are resolvable in mass.

\begin{figure}[t]
\begin{center}
\includegraphics[scale=0.4]{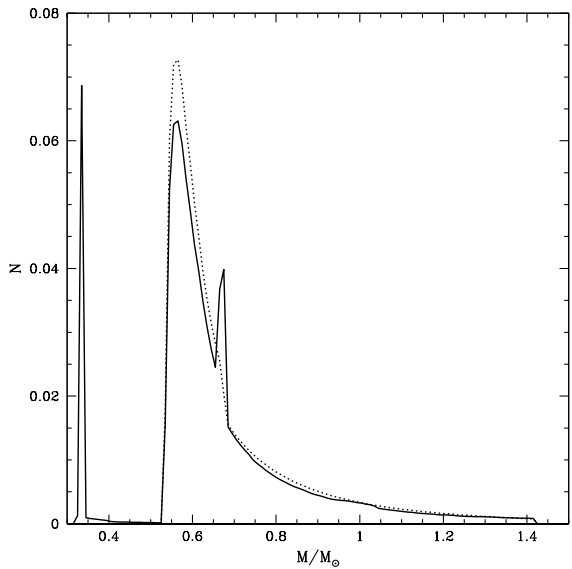}
\caption{Normalized mass distributions  of single white dwarfs (dotted
  line) and single plus binary white dwarfs (solid line) assuming that
  all the stars belonging to binary systems have known masses.}
\label{fig1}
\end{center}
\end{figure}

The  stellar  data for  the  white  dwarf  progenitors is  taken  from
\cite{domi99}  and the adopted  relationship between  the mass  of the
progenitor  and the  mass  of the  resulting  white dwarf  is that  of
\cite{cata08}. The  cooling times are from \cite{sere02}  for He white
dwarfs,  from  \cite{sala00,sala10}  for  CO  white  dwarfs  and  from
\cite{alth07} for  the ONe white  dwarfs. In these calculations  it is
assumed that after  merging, the new white dwarf  has the same initial
temperature and chemical composition as  white dwarfs of the same mass
born from a single star.

\section{Results and discussion}

\begin{figure}
\begin{center}
\includegraphics[scale=0.4]{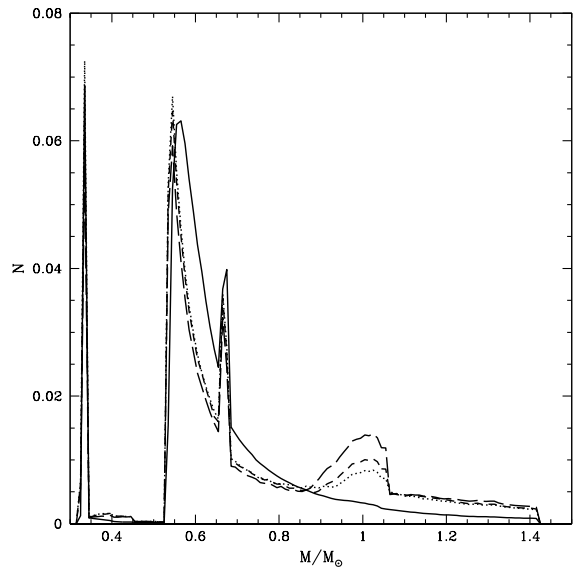}
\caption{Mass distributions  of cold and hot binary  white dwarfs. The
  solid  line represents  the mass  distribution of  all  white dwarfs
  while the broken ones those of hot white dwarfs. White dwarfs hotter
  than 15,000, 13,000 and 12,000 K are represented by the long-dashed,
  dashed and dotted lines respectively.}
\label{fig2}
\end{center}
\end{figure}

Fig.~\ref{fig1} displays how the mass function of white dwarfs changes
when the evolution of binaries is included (solid line) or not (dotted
line).  The  first noticeable difference  is the presence of  a narrow
spike at  $\sim 0.30\, M_{\sun}$, which  is entirely due to the presence of He
white dwarfs in binary systems. This is a well  known result and its extreme thinness is
probably caused  by the  simplicity of the  model used here.  The main
peak of the distribution is smaller than in the case in which binaries
are not  taken into account due to  the fact that in  many binaries CO
white dwarfs cannot form. The  secondary peak appearing at $\sim 0.7\,
M_{\sun}$ is  caused by  the merger of  two He  white dwarfs and  is a
consequence of the assumption that these objects evolve  to form a  CO white
dwarf.  Therefore, the properties of this peak, including its absence,
could provide important  insights into the process of  merging of such
objects.   The  region  above  this  peak  is  rather  smooth  and  is
characterized by the  fact that for $M_{\rm WD}  \la 1\, M_{\sun}$ the
number of white dwarfs in the case in which binaries are considered is
systematically smaller than in the  simulation in which are not, while
for $M_{\rm  WD} \ga 1\,  M_{\sun}$ this is  not case. This  effect is
caused by mergers, but its influence is small (at $1\, M_{\sun}$ CO+He
mergers represent $\sim 17.7\%$ of  all white dwarfs and CO+CO mergers
$\sim 5.5\%$), so there is not any prominent bump.

In order to compare with the observations it is necessary to take into
account that the  mass distribution is only known  for DA white dwarfs
with $T_{\rm  eff} \ga 12\,000 - 13\,000$  K. This is due  to the fact
that hotter  white dwarfs can  contain important amounts of  helium in
their  atmospheres,  which  makes  the  determination  of  their  mass
unreliable  \citep{berg92,giam12}.  Fig.~\ref{fig2} compares  the mass
distribution function  of hot white  dwarfs (dashed and  dotted lines)
with  that of  all white  dwarfs  (solid line).   The most  noticeable
feature is the prominet bump around  $1 \, M_{\sun}$. It is clear that
this behavior  is not caused by  the mergers but by  the dependence of
the cooling rates with the mass. Moreover, this bump is not present in
the mass distribution obtained from the SDSS catalogue, but it appears
in the mass distribution of nearby stars.

\section{Conclusions}

Our toy model  indicates that effectively the interaction  of stars in
close  enough binary systems  can introduce  important changes  in the
expected  mass  distribution  of   white  dwarfs  resulting  from  the
evolution of single  stars. Besides the existence of  He white dwarfs,
the merging of such stars can produce a pronounced bump around $0.7-0.8
\, M_{\sun}$. The exact location and shape of this bump depends on the
details of the merging process. The merger  of CO and He and of two CO
white   dwarfs  only   introduces   small  variations   in  the   mass
distribution. In the case of hot  white dwarfs we have found a bump at
$\sim\, 1  M_{\sun}$ caused by the  dependence of the  cooling rate on
the mass of the white dwarf.  This bump is absent in the mass function
obtained from  the SDSS catalogues  but there is  a hint of it  in the
mass function of the local sample.

\acknowledgements 
This  research   has  been   partially  supported  by   MICINN  grants
AYA2008--04211--C02--01  and  AYA2011--24704,  by  the  ESF  EUROGENESIS
project  (MICINN grant  EUI2009--04167), by  the European  Union FEDER
funds and by the AGAUR.

\bibliography{isern}

\end{document}